\documentclass{jfm}

\usepackage{graphicx}
\usepackage{natbib}
\usepackage{bm} % <<<< NEW
\usepackage{color} % <<<< NEW

\usepackage{amsmath}

\ifCUPmtlplainloaded \else
  \checkfont{eurm10}
  \iffontfound
    \IfFileExists{upmath.sty}
      {\typeout{^^JFound AMS Euler Roman fonts on the system,
                   using the 'upmath' package.^^J}%
       \usepackage{upmath}}
      {\typeout{^^JFound AMS Euler Roman fonts on the system, but you
                   dont seem to have the}%
       \typeout{'upmath' package installed. JFM.cls can take advantage
                 of these fonts,^^Jif you use 'upmath' package.^^J}%
      }
  \else
  \fi
\fi

\ifCUPmtlplainloaded \else
  \checkfont{msam10}
  \iffontfound
    \IfFileExists{amssymb.sty}
      {\typeout{^^JFound AMS Symbol fonts on the system, using the
                'amssymb' package.^^J}%
       \usepackage{amssymb}%
       \let\le=\leqslant  
       \let\ge=\geqslant  
      }{}
  \fi
\fi

\ifCUPmtlplainloaded \else
  \IfFileExists{amsbsy.sty}
    {\typeout{^^JFound the 'amsbsy' package on the system, using it.^^J}%
     \usepackage{amsbsy}}
    {\providecommand\boldsymbol[1]{\mbox{\boldmath $##1$}}}
\fi

\newsavebox{\astrutbox}
\sbox{\astrutbox}{\rule[-5pt]{0pt}{20pt}}

\title[Recursive DMD]{Recursive dynamic mode decomposition of
a transient cylinder wake}

\author[B.~R.~Noack,
        W.~Stankiewicz,
        M.~Morzy\'nski and
        P.~J.~Schmid]%
{Bernd R. Noack$^{1,2}$%
  \thanks{Email address for correspondence: bernd.noack@univ-poitiers.fr},\ns
Witold Stankiewicz$^3$\break
Marek Morzy\'nski$^3$ and Peter J. Schmid$^4$}

\affiliation{
  $^1$Institut PPRIME, CNRS - Universit\'e de Poitiers - ENSMA, UPR
  3346, D\'epartment Fluides, Thermique, Combustion, CEAT,
  43 rue de l'A\'erodrome, F-86036 POITIERS Cedex, France\\[\affilskip]
  $^2$Institut f\"ur Str\"omungsmechanik, Technische Universit\"at
  Berlin,
  Hermann-Blenk-Stra{\ss}e 37, D-38108 Braunschweig, Germany\\[\affilskip]
  $^3$Division of Virtual Engineering, Institute of Combustion Engines
  and Transport, Pozna\'n University of Technology,
  Piotrowo 3 street, 60-965 Pozna\'n, Poland\\[\affilskip]
  $^4$Department of Mathematics, Imperial College London, South
  Kensington Campus, London SW7 2AZ, United Kingdom}

\pubyear{2014}
\volume{650}
\pagerange{119--126}
\date{2015-11-21}
\begin{document}

\maketitle

%--------------------------------------------------------------------------
\begin{abstract}
  A novel data-driven modal decomposition of fluid flow is proposed
  comprising key features of POD and DMD. The first mode is the
  normalized real or imaginary part of the DMD mode which minimizes
  the time-averaged residual. The $N$-th mode is defined recursively
  in an analogous manner based on the residual of an expansion using
  the first $N-1$ modes. The resulting recursive DMD (RDMD) modes are
  orthogonal by construction, retain pure frequency content and aim at
  low residual. RDMD is applied to transient cylinder wake data and is
  benchmarked against POD and optimized DMD \citep{Chen2012jns} for
  the same snapshot sequence. Unlike POD modes, RDMD structures are
  shown to have pure frequency content while retaining a residual of
  comparable order as POD. In contrast to DMD with exponentially
  growing or decaying oscillatory amplitudes, RDMD clearly identifies
  initial, maximum and final fluctuation levels. Intriguingly, RDMD
  outperforms both POD and DMD in the limit cycle resolution from the
  same snaphots. RDMD is proposed as an attractive alternative to POD
  and DMD for empirical Galerkin models, with nonlinear transient
  dynamics as a niche application.
\end{abstract}

%--------------------------------------------------------------------------
%\begin{keywords}
%  Authors should not enter keywords on the manuscript, as these must
%  be chosen by the author during the online submission process and
%  will then be added during the typesetting process (see
%  http://journals.cambridge.org/data/\linebreak[3]relatedlink/jfm-\linebreak[3]keywords.pdf
%  for the full list)
%\end{keywords}
%--------------------------------------------------------------------------

%\tableofcontents
%--------------------------------------------------------------------------
\section{Introduction}
\label{ToC:Introduction}
%--------------------------------------------------------------------------

This study proposes a novel flow field expansion tailored
to the construction of 
low-dimensional empirical Galerkin models. Such reduced-order models
(1) help in data compression, (2) allow quick visualizations and
kinematic mixing studies \citep{RomKedar1990jfm}, (3) provide a
testbed for physical understanding \citep{Lorenz1963jas}, (4) serve as
computationally inexpensive surrogate models for optimization
\citep{Han2013ast} or (5) may be used as a plant for control design
\citep{Gerhard2003aiaa,Bergmann2008jcp}.

In 1858, Helmholtz has laid the foundation for the first
low-dimensional dynamical models in fluid mechanics with his famous
theorems on vortices \citep[see, e.g.][]{Lugt1995book}. Subsequently,
a rich set of vortex models have been developed for vortex pairs, for
the recirculation zone \citep{Foeppl1913,Suh1993jpsj}, for the vortex
street \citep{Karman1912}, and for numerous combustion related
problems \citep{Coats1997pecs}, to name just a few. For non-periodic
open flows, low-dimensional vortex models come at the expense of a
hybrid state-space structure: new degrees of freedom (vortices) are
created at the body, merged or removed.  Most forms of applications,
like dynamical systems analyses or control design, are not suitable
for hybrid models but assume a continuous evolution in a fixed
finite-dimensional state space.  Hence, most currently developed
reduced-order models are formulated by the Galerkin method and based
on modal expansions \citep{Fletcher1984book}.

In the last 100 years \citet{Galerkin1915vi}'s method of solving
partial differential equations has enjoyed ample generalisations and
applications, from high-dimensional grid-based computational methods
to low-dimensional models. This study focusses on low-dimensional flow
representations by an expansion in global modal structures. In
principle, any space of square-integrable velocity fields has a
complete set of orthonormal modes: any flow field can be arbitrarily
closely approximated by a finite-dimensional expansion. In practice,
however, the construction of such mathematical bases is restricted to
simple geometries and the use of Fourier expansions or Chebyshev
polynomials \citep{Orszag1971jfm}. Stability modes based on a
linearisation of the Navier-Stokes equation tend to be more efficient
in terms of resolution for a given number $N$ of modes. However, these
physical modes generally lack a proof of completeness, and are
afflicted by a reduced dynamic bandwidth and an ${\cal{O}}(N^3)$
operation count for the quadratic terms as opposed to ${\cal{O}}(N
\log N)$ operations for Fourier or Chebyshev modes. The most efficient
representations of a Navier-Stokes solution are obtained from
empirical expansions based on flow snapshots. These data-driven
Galerkin expansions are confined to a subspace spanned by the
snapshots, i.e.\ have a narrow dynamic bandwidth defined by the
training set.

This study focusses on data-driven expansions. A Galerkin expansion
with guaranteed minimal residual over the training snapshots was first
pioneered by \citet{Lorenz1956mit} as principal axis modes and later
popularized in fluid mechanics as proper orthogonal decomposition
(POD) by \citet{Lumley1967proc}. POD guarantees an optimal data
reconstruction in a well-defined sense \citep[see,
e.g.,][]{Holmes1998book}. Apart from data compression applications,
Lumley devised POD as a mathematical tool for distilling coherent
structures from data. Yet, only in rare cases have POD modes been
found to resemble physically meaningful structures, like stability
modes or base-flow deformations \citep{Oberleithner2011jfm}. In
particular, they tend to mix spatial and temporal frequencies in most
modes which complicates a physical interpretation. As a remedy for
this shortcoming, the dynamic mode decomposition (DMD) --- also
referred to as Koopman analysis --- was pioneered by
\citet{Rowley2009jfm} and \citet{Schmid2010jfm}. DMD is able to
distill stability eigenmodes from transient snapshot data and produce
temporal Fourier modes for post-transient data. The downside of these
DMD features --- as compared to POD --- are non-orthogonality of the
extracted modes, suboptimal convergence, the need for time-resolved
snapshots and numerical challenges when filtering is omitted. Over the
past, numerous Galerkin models based on POD and DMD have been
constructed. In addition, numerous generalisations have been proposed
to address, among other topics, multi-operating conditions
\citep{Jorgensen2003tcfd}, changes during transients
\citep{Siegel2008jfm}, optimal correlations to observables
\citep{Hoarau2006pf,Schlegel2012jfm}, and control design
\citep{Brunton2015amr}.

This study proposes a novel data-driven expansion preserving key
features of POD, such as orthonormality of the computed modes and a
low residual, and of DMD, such as distilling the dominant frequencies
and their associated structures contained in the data. The manuscript
is organised as follows. \S~\ref{ToC:DNS} describes the cylinder wake
configuration and gives details on the employed direct numerical
simulation as well as the extracted snapshots. \S~\ref{ToC:Method}
outlines the computation of the proposed expansion, called 'Recursive
DMD (RDMD)' in what follows. RDMD is then applied to a transient
cylinder wake, demonstrating its advantages (\S~\ref{ToC:Results})
over previous decompositions. Our results are summarised in
\S~\ref{ToC:Conclusions}.

%--------------------------------------------------------------------------
\section{Configuration and direct numerical simulation}
\label{ToC:DNS}
%--------------------------------------------------------------------------

As a test-case, a two-dimensional, incompressible, viscous flow past a
circular cylinder has been chosen. The flow is described in a
Cartesian coordinate system where the $x$-axis is aligned with
streamwise direction and the $y$-axis is transverse and orthogonal to
the cylinder axis. The origin of the coordinate system coincides with
the cylinder axis. The location vector is denoted by $\bm{x}=(x,y)$.
Analogously, the velocity is represented by $\bm{u}= (u,v)$ where $u$
and $v$ are the $x$- and $y$-components of the velocity
components. The time is presented by $t$. The Newtonian fluid is
characterised by the density $\rho$ and dynamic viscosity $\mu$. In
the following, all variables are assumed to be nondimensionalised by
the cylinder's diameter $D$, the oncoming velocity $U$ and the fluid
density $\rho$. The resulting Reynolds number $Re=U D \rho /\mu$ is
set to $100$, i.e.\ well above the onset of vortex shedding
\citep{Zebib1987jem,Schumm1994jfm}, but well below the onset of
three-dimensional instabilities
\citep{Zhang1995pf,Barkley1996jfm,Williamson1996arfm}.

A direct numerical simulation of the Navier-Stokes equations has been
performed using an in-house solver based on a second-order
finite-element discretisation with Taylor-Hood elements
\citep{Taylor1973cf} in penalty formulation. The time stepping is
performed using a third-order accurate scheme and a time step equal to $0.1$. Following
\citet{Noack2003jfm}, the computational domain extends from $x=-5$ to
$x=25$ and $y=-5$ to $y=5$ and is discretised with 56272 finite
elements.

\begin{figure}
  \centering{ %
    \begin{tabular}{cc}
      {\includegraphics[width=0.4\textwidth]{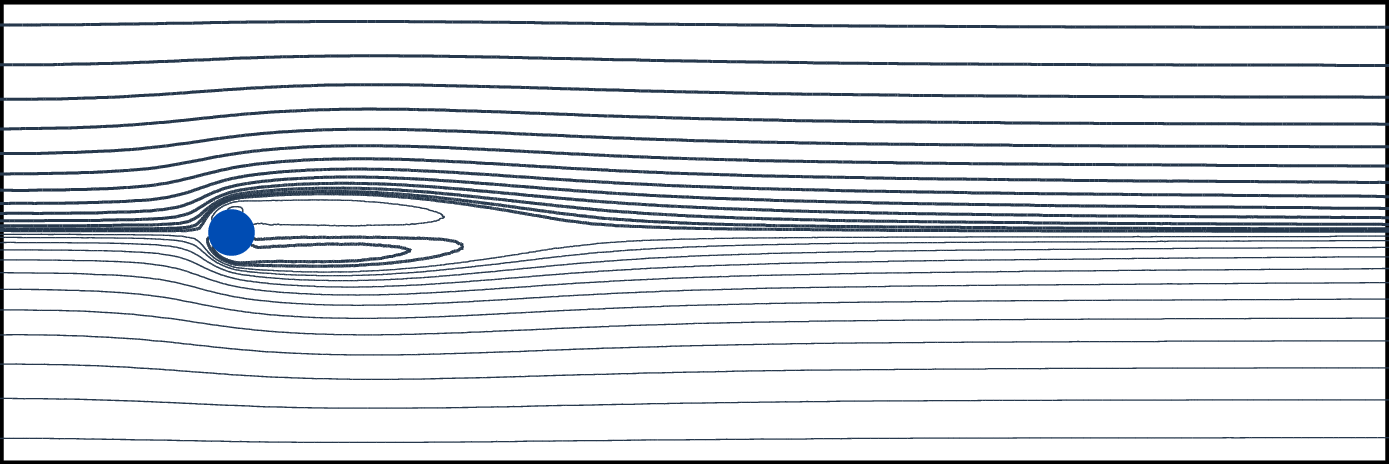}}  &
      {\includegraphics[width=0.4\textwidth]{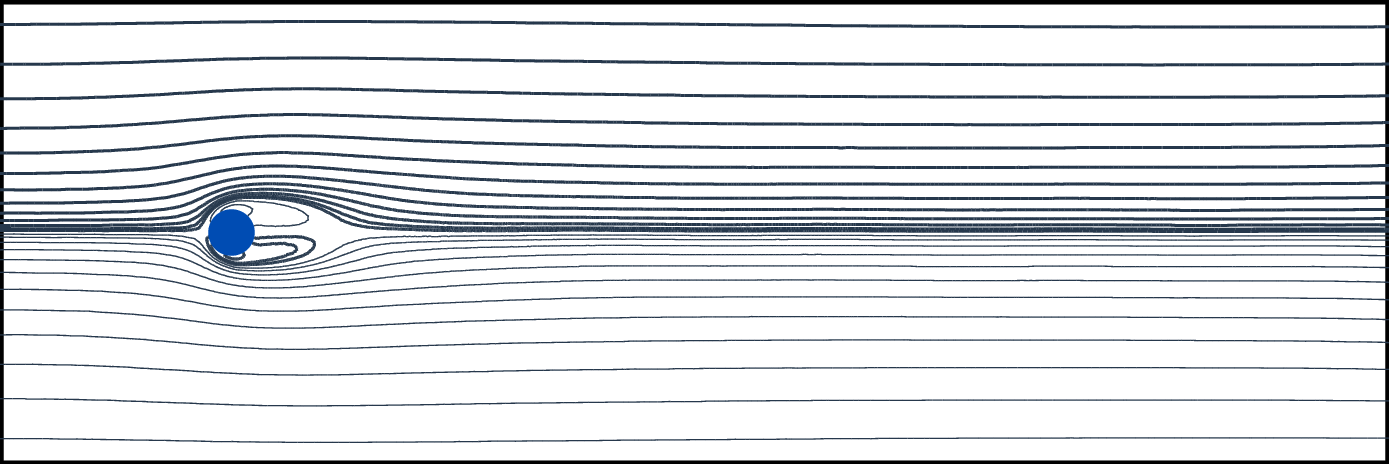}} \tabularnewline
      \multicolumn{2}{c}{
        \includegraphics[width=0.8\textwidth]{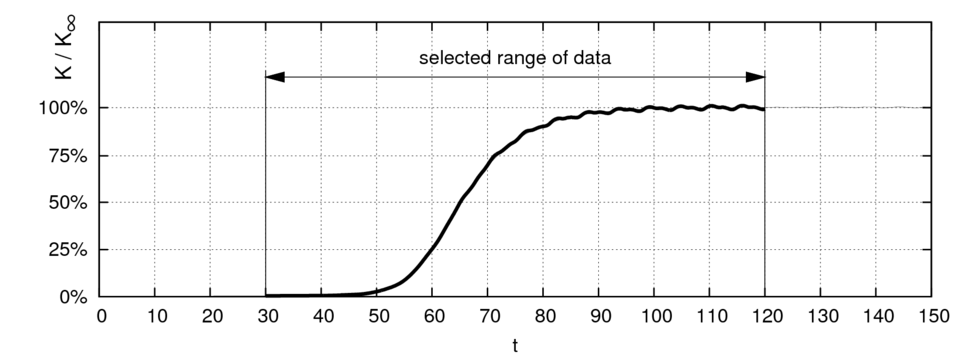}
      }\tabularnewline
    \end{tabular}
    \caption{Incompressible two-dimensional flow around a circular
      cylinder at $Re=100$. The steady (top-left) and time-averaged
      (top-right) solutions are illustrated with streamlines. The
      total fluctuation level normalized with its asymptotic value is monitored below.}
    \label{fig:base_cylinderfine} }
\end{figure}

The simulation provides $M=450$ equidistantly sampled velocity
snapshots $\bm{u}^m(\bm{x}) = \bm{u}(\bm{x},t^m)$, $m=1,..,M$,
covering the entire unforced transient phase, from the steady solution
to the fully-developed von K\'arm\'an vortex street. The snapshot
times are $t^m = 30 + m \Delta t $ based on a sampling interval $\Delta t =
0.2$ and cover the time interval $t \in [30,120].$ The initial and
final base flows are depicted in figure \ref{fig:base_cylinderfine}
together with the energy norm of the fluctuations $\bm{v} =
\bm{u}-\bm{u}_s$ around the steady solution $\bm{u}_s$. The transition
from the unstable fixed point to the stable limit-cycle oscillation is
characterised by the growing amplitudes of the fluctuations
accompanied by an increase in the Strouhal number
\citep{Zebib1987jem,Schumm1994jfm}.

%----- Figure ----------------------------------------------------------
%----- End of figure ---------------------------------------------------
%%%% ACTION: The y-caption should be in capital letters.

\begin{figure}
  \centering{ %
    {\includegraphics[width=1.0\textwidth]{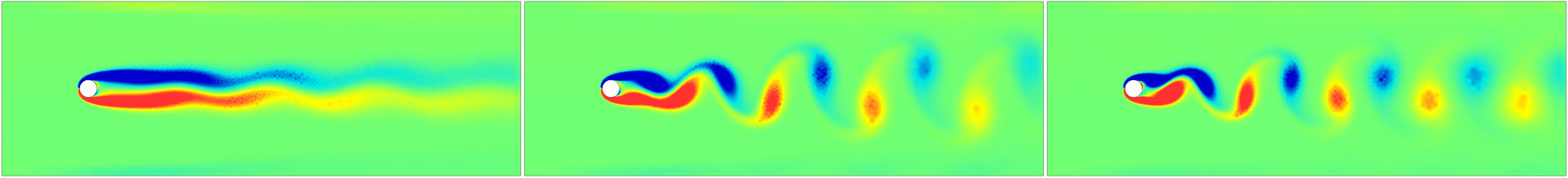}}
  }
  \caption{Transient evolution of the cylinder wake illustrated with
    snapshots of the vorticity field at an initial (left, $t=50$), an
    intermediate (middle, $t=70$) and final state (right, $t=90$).
    \label{fig:flow-transition}}
\end{figure}

Within the first 30 convective time units, the flow is governed by
linear dynamics spanned by the steady solution and the unstable
eigenmode. In the intermediate phase, approximately $t \in [60,90]$,
the vortex shedding undergoes significant chances and moves farther
upstream towards the cylinder. In the final 30 convective time units,
the flow has converged to a limit-cycle dynamics exhibiting the
fully-developed von K\'arm\'an vortex street. These three
characteristic stages of the transient evolution are depicted in
figure \ref{fig:flow-transition}.

%----- Figure ----------------------------------------------------------
%----- End of figure ---------------------------------------------------

All modal decompositions are based on these fluctuations around the
steady solution.  The mean flow is discarded as a base flow because it
is only well defined for this particular initial condition and the
chosen time interval. Our sampling Strouahl frequency of $10$ 
is about 30 times larger than the shedding frequency -- a value
which can be considered adequate for a Fourier transformation while
avoiding excessive redundancy for the statistical POD.
%%% ACTION: Note that I have changed the Delta t = 0.2 (!!!)
%**************************************************************************
\section{Modal decomposition}
\label{ToC:Method}

In this section, a new snapshot-based modal decomposition is proposed.
This decomposition comprises properties of the Proper Orthogonal
Decomposition (POD) \citep{Lumley1967proc,Sirovich1987qam1} and the
Dynamic Mode Decomposition (DMD) \citep{Rowley2009jfm,Schmid2010jfm}.
First (\S~\ref{ToC:Method:POD} \& \ref{ToC:Method:DMD}), the
snapshot-based POD and DMD are briefly recapitulated. In
\S~\ref{ToC:Method:RDMD}, the recursive DMD (RDMD) is proposed as an
appealing compromise inheriting the orthonormality and low truncation error of
POD and the oscillatory representation of the flow behaviour of DMD.

%--------------------------------------------------------------------------
\subsection{Proper Orthogonal Decomposition}
\label{ToC:Method:POD}

We analyse a time-dependent velocity field $\bm{u}(\bm{x},t)$ in a
steady domain $\bm{x} \in \Omega$ and sample $M$ flow snapshots with
constant sampling frequency corresponding to a time step $\Delta t$.
The velocity field at instant $t^m = m \Delta t$, with $m=1,\ldots,
M,$ is denoted by $\bm{u}^m := \bm{u}(\bm{x},t^m)$.

POD requires the definition of an inner product and a time average.
We assume the standard inner product of two square-integrable velocity
fields $\bm{u}$, $\bm{v} \in {\cal L}^2 (\Omega),$ given as
%-----------------------------------------------------------------------
\begin{equation}
  \left ( \bm{u}, \bm{v} \right )_{\Omega}
  := \int\limits_{\Omega} d\bm{x} \> \bm{u}(\bm{x}) \cdot \bm{v}(\bm{x}).
\end{equation}
%-----------------------------------------------------------------------
The corresponding norm reads $\Vert \bm{u} \Vert_{\Omega} :=
\sqrt{\left ( \bm{u}, \bm{u} \right )_{\Omega}}$.  The snapshot-based
time average of a function $F$ is defined in a canonical manner:
\begin{equation}
  \langle F(\bm{u})  \rangle_M :=
  \frac{1}{M} \sum\limits_{m=1}^{M}  F \left( \bm{u}^m \right).
\end{equation}

In the following, the snapshot-based POD \citep{Sirovich1987qam1} is
applied to the fluctations 
{$\bm{v}^m (\bm{x}) :=\bm{v}(\bm{x},t^m)-\bm{u}_s (\bm{x})$
around the steady solution $\bm{u}_s$ 
for the reasons mentioned in \S~\ref{ToC:DNS}.
%%% ACTION: Note that I have removed the mean flow.
First, the correlation matrix $\bm{C} = \left( C^{mn} \right) \in {\Bbb{R}}^{M \times M}$
of the fluctuations is determined,
\begin{equation}
  \label{eqn:Cmn}
  C^{mn} := \frac{1}{M}
  \left ( \bm{v}^m , \bm{v}^n  \right )_{\Omega}.
\end{equation}
Second, a spectral analysis of this matrix is performed. Note that
$\bm{C}$ is a symmetric, positive semi-definite Gramian matrix.
Hence, its eigenvalues $\lambda_i$ are real and nonnegative and can be
assumed to be sorted according to $\lambda_1 \ge \lambda_2 \ge \ldots
\ge \lambda_M \ge 0$. The corresponding eigenvectors $\bm{e}_i =
\left[ e_i^1, \ldots, e_i^M\right]^T$ satisfy
\begin{equation}
  \label{eqn:Cmn:ev}
  \bm{C} \bm{e}_i = \lambda_i \bm{e}_i, \qquad i=1, \ldots, M.
\end{equation}
and can -- without loss of generality -- be assumed to be
orthonormalized, satisfying $\bm{e}_i \cdot \bm{e}_j = \delta_{ij}$
with $\delta_{ij}$ as the Kronecker symbol. Third, each POD mode is
expressed as a linear combination of the snapshot fluctuations,
\begin{equation}
  \label{eqn:pod:mode}
  \bm{u}_i :=
  \frac{1}{\sqrt{M \> \lambda_i}} \>
  \sum\limits_{m=1}^{M} e_i^m \>
  \bm{v}^m , \qquad i=1, \ldots, N.
\end{equation}
It follows that the POD modes are orthonormal, with
$\left ( \bm{u}_i, \bm{u}_j \right)_{\Omega} = \delta_{ij},
 \quad \forall i,j\in \{1,\ldots,M\}$.
Finally, the mode amplitudes read
\begin{equation}
\label{eqn:pod:amplitudes}
a_i^m :=  \sqrt{\lambda_i \> M} \> e^{i}_m, \qquad i=1, \ldots, N.
\end{equation}
These amplitudes  are uncorrelated (orthogonal in
time), or in mathematical terms
\begin{equation}
  \label{eqn:a_i:moments}
  \langle a_i a_j \rangle_M = \lambda_i \> \delta_{ij}, \qquad
  i,j \in \{1, \ldots, N\}.
\end{equation}
Note that the first moments $\langle a_i \rangle_M$
do not need to vanish as the fluctuations are based on the steady solution
and not on the mean flow of this transient.
The POD defines a second-order statistics providing the two-point autocorrelation function
\begin{equation}
  \label{eqn:pod:correlation}
  \overline{
    \bm{v} (\bm{x},t) \>
    \bm{v} (\bm{y}, t) }
  = \sum\limits_{i=1}^N \lambda_i \>
  \bm{u}_i(\bm{x}) \> \bm{u}_i(\bm{y}).
\end{equation}
Hence, a minimum requirement imposed on the snapshot ensemble is the
accuracy of the extracted mean flow and the flow's second moments.
This accuracy of the statistics for a given number of snapshots is
increased by processing uncorrelated snapshots as required in the
original paper on the snapshot POD method \citep{Sirovich1987qam1}.

The resulting expansion exactly reproduces the snapshots for $N=M$
modes; we have
\begin{equation}
  \label{Eqn:GA}
  \bm{u} (\bm{x},t^m) = \bm{u}_s(\bm{x})
  + \sum\limits_{i=1}^N a_i(t^m) \> \bm{u}_i(\bm{x}).
\end{equation}
%%%% ACTION: Note that N=M not N=M-1 for u_s!
%%%% I have added the definition of the residual level.
For $N < M$, the truncated expansion \eqref{Eqn:GA}
has a non-vanishing residual $\bm{r}^m(\bm{x}) := \bm{r}(\bm{x},t^m)$.
The corresponding time-averaged truncation error 
\begin{equation}
\label{Eqn:TruncationError} 
\chi^2 := \langle \Vert \bm{r}^m \Vert_{\Omega}^2\rangle_M
\end{equation}
can be shown to be minimal; in other words, no other Galerkin expansion will
have a smaller error
\citep{Holmes1998book}. 
This optimality property makes POD an
attractive data compression technique.

For later reference, the instantaneous truncation error 
$\chi^2(t) :=   \Vert \bm{r}^m(\cdot,t) \Vert_{\Omega}^2$ is introduced.
The size of this error may be compared 
with the corresponding fluctuation level on the limit cycle
$2K_{\infty} = \overline{\Vert \bm{v} \Vert^2}$,
were $K_{\infty}$ denotes  the turbulent kinetic energy (TKE)
and the overbar represents averaging over the post-transient phase.
We also introduce $K$ as the corresponding instantaneous quantity.

As a motivation for the proposed new decomposition, we recall that POD
can also be defined in a recursive manner, following
\citet{Courant1989book} on the spectral analysis of positive definite
symmetric matrices. Taking $\bm{u}_1$ as the first expansion mode,
the resulting one-mode expansion reads
\begin{equation}
  \bm{v}^m = a_1^m \bm{u}_1 + \bm{r}_1^m, \quad m=1,\ldots,M .
\end{equation}
The mode amplitude $a_1^m := (\bm{v}^m, \bm{u}_1)_{\Omega}$ minimizes
the residual $\Vert \bm{r}_1^m \Vert_{\Omega}$ for a given $\bm{u}_1$.
The first POD mode can be shown to minimize the averaged energy of the
residual $\langle \Vert \bm{r}_1^m \Vert_{\Omega}^2 \rangle_M $.
Furthermore, the residual $\bm{r}_1^m$, $m=1,\ldots,M$ is orthogonal
to $\bm{u}_1$ by construction.  The second step then searches for a
mode $\bm{u}_2$ which best resolves the residual $\bm{r}_1$,
\begin{equation}
  \bm{r}^m_1 = a_2^m \bm{u}_2 + \bm{r}_2^m  \quad m=1,\ldots,M ,
\end{equation}
i.e.\ which minimizes $\left \langle \Vert \bm{r}_2^m \Vert_{\Omega}^2
\right \rangle_M$.  The other remaining modes are computed in a
similar manner.

%--------------------------------------------------------------------------
\subsection{Dynamic Mode Decomposition}
\label{ToC:Method:DMD}

The Dynamic Mode Decomposition \citep{Rowley2009jfm,Schmid2010jfm} is
another data-driven modal expansion which can approximate stability
eigenmodes from transient data or Fourier modes from post-transient
data. The time step $\Delta t$ needs to be sufficiently small for a
meaningful Fourier analysis, but sufficiently large so that the
changes in the flow state exceed the noise level.

We consider the fluctuation snaphots of \S~\ref{ToC:Method:POD},
$\bm{v}^m$, $m=1, \ldots, M $ as linearly independent modes of an
expansion and write
\begin{equation}
  \label{Eqn:DMD:v}
  \bm{v} = \sum\limits_{i=1}^M  b_i \bm{v}^i.
\end{equation}
Evidently, the modes are generally not orthogonal. With this basis,
the mode amplitude vector of the $m$-th snapshot becomes a unit
vector:
\begin{equation}
  \label{Eqn:DMD:b}
  \bm{b}^m = \left[ b_1^m, \ldots, b_M^m \right]^T \quad
  \hbox{where} \quad
  b_i^m = \delta_{im}.
\end{equation}
DMD assumes a linear relationship between the $(m+1)$-th and $m$-th
snapshot
\begin{equation}
  \label{Eqn:DMD:bnext}
  \bm{b}^{m+1} = \bm{A} \> \bm{b}^m,
\end{equation}
where $\bm{A} \in {\Bbb{R}}^{M \times M}$ is a square matrix which is
generally identified from the data. From Eqn.\ \eqref{Eqn:DMD:bnext},
the matrix is easily seen to be
\begin{equation}
  \label{Eqn:DMD:A}
  \bm{A} = \left[
    \begin{array}{cccccc}
      0 & 0 & \hdots & 0 & 0 & c_1 \\
      1 & 0 & \hdots & 0 & 0 & c_2 \\
      \vdots & \vdots & \ddots & \vdots & \vdots \\
      0 & 0 & \hdots & 1 & 0 & c_{M-1} \\
      0 & 0 & \hdots & 0 & 1 & 0
    \end{array}
  \right].
\end{equation}
For $i<M$ the matrix acts as a shift map for the only non-vanishing
element of $\bm{b}$. The last snaphot is expanded in terms of the
previous ones:
\begin{equation}
  \bm{v}^M = \sum\limits_{i=1}^{M-1} c_i \> \bm{v}^i + \bm{r}.
\end{equation}
The coefficients $c_i$ are chosen to minimize the residual norm $\Vert
\bm{r} \Vert_{\Omega}$.

In what follows, we depart from the classical DMD literature and
propose a simpler derivation of the DMD modes. Let $P(s) = -c_1 - c_2
s - \ldots -c_{M-1} s^{M-2} - s^M$ be a polynomial in $s$ and let $
P(s) = (s-\lambda_1) (s-\lambda_2) \ldots (s-\lambda_M)$ be its
factorization with distinct eigenvalues $\lambda_i$. We introduce
\begin{equation}
  \label{Eqn:DMD:V}
  \bm{V} = \left[
    \begin{array}{cccc}
      1 & \lambda_1 & \hdots & \lambda_1^{M-1} \\
      1 & \lambda_2 & \hdots & \lambda_2^{M-1} \\
      \vdots & \vdots & \vdots & \vdots \\
      1 & \lambda_M & \hdots & \lambda_M^{M-1}
    \end{array}
  \right].
\end{equation}
as the corresponding Vandermonde matrix. It can be easily verified
that the Vandermonde matrix $\bm{V}$ diagonalizes the companion matrix
$\bm{A}.$ We obtain
\begin{equation}
  \label{Eqn:DMD:A:Diagonalisation}
  \bm{V} \bm{A} \bm{V}^1 = \hbox{diag}(\lambda_1,\ldots,\lambda_M),
\end{equation}
where the right-hand side is a diagonal matrix with the eigenvalues as
its elements. We introduce new variables $\bm{a}$ defined by
\begin{equation}
  \label{Eqn:DMD:aDefinition}
  \bm{a} = \bm{V} \bm{b}.
\end{equation}

With these definitions, the evolution equation \eqref{Eqn:DMD:bnext}
can be cast into eigenform according to
\begin{equation}
  \label{Eqn:DMD:anext}
  \bm{a}^{m+1} = \bm{V} \bm{C} \bm{V}^{-1} \bm{a}^m = \bf{D} \bm{a}^m.
\end{equation}
Here, $\bm{D}$ denotes the diagonal matrix
\begin{equation}
  \label{Eqn:DMD:D}
  \bm{D} = \hbox{diag} (\lambda_1,\ldots,\lambda_M)
  = \left[
    \begin{array}{cccc}
      \lambda_1 & 0         & \hdots & 0 \\
      0         & \lambda_2 & \hdots & 0 \\
      \vdots    & \vdots    & \vdots & \vdots \\
      0         &         0 & \hdots & \lambda_M
    \end{array}
  \right]
\end{equation}
where the $i$-th eigenvalue is $\lambda_i$ with corresponding
eigenvector $\bm{a}_i = \left[\delta_{1i},\ldots,\delta_{Mi}
\right]^{T}.$

Equations \eqref{Eqn:DMD:b}, \eqref{Eqn:DMD:bnext} and
\eqref{Eqn:DMD:aDefinition} imply that the snapshots can be expressed
as
\begin{equation}
  \label{Eqn:DMD:Constitutive}
  \bm{v}^m = \sum\limits_{i=1}^{M-1} \lambda_i^{m-1} \bm{\Phi}_i
\end{equation}
where $\lambda_i$ are referred to as DMD eigenvalues and $\bm{\Phi}_i$
as (complex) DMD modes. Note that the derivation of
\eqref{Eqn:DMD:Constitutive} rests on the diagonalization of the
companion matrix \eqref{Eqn:DMD:A:Diagonalisation} and does not
require the Koopman operator.

The choice of  $N$ DMD modes 
for the Galerkin expansion \eqref{Eqn:GA}
minimizes the truncation error \eqref{Eqn:TruncationError}
following \citet{Chen2012jns}
and consistent with the optimal property of POD.
%%% ACTION: This sentence has been added today 2015-11-18
%%% and was necessary.

%--------------------------------------------------------------------------
\subsection{Recursive Dynamic Mode Decomposition}
\label{ToC:Method:RDMD}

The recursive Dynamic Mode Decomposition (RDMD) serves a
multi-objective task: extracting oscillatory modes from the snapshot
sequence (like DMD) while ensuring orthogonality of the modes and a
low truncation error (like POD).

The initialisation step prepares the residual to be processed. We take
\begin{equation}
  \label{Eqn:RDMD1}
  \bm{r}_0^m := \bm{v}^m, \qquad m=1,\ldots,M.
\end{equation}

During the $i$-th step ($0<i \le N$), the $i$-th mode is determined
from a DMD
\begin{equation}
  \label{Eqn:RDMD2}
  \bm{r}_{i-1}^m = \sum\limits_{j=1}^{M-1} \lambda_j^m \> \bm{\Phi}_j.
\end{equation}
%%% ACTION: Note that I have changed index bounds to be consistent with (3.22)
%%%  \label{Eqn:DMD:Constitutive}
The candidate modes to be considered are
\begin{equation}
  \label{Eqn:RDMD3}
  \bm{u}_{n}^\star = \frac{ \Re\{ \bm{\Phi}_n \} }{
    \left \Vert \Re\{ \bm{\Phi}_n \} \right \Vert_{\Omega} },
\end{equation}
and each mode reduces the snapshot-dependent residual to
$\bm{r}_{i}^m$ according to
\begin{subequations}
  \begin{eqnarray}
    \label{Eqn:DMD4}
    \bm{r}_{i-1}^m &=& a_n^m \bm{u}_n^{\star} + \bm{r}_{i}^m, \\
    a_n^m &=& \left ( \bm{r}_{i-1}^m, \bm{u}_n^{\star} \right)_{\Omega}.
  \end{eqnarray}
\end{subequations}
The truncation error of the $i$-th candidate mode $\bm{u}_n^{\star}$
for all snapshots is given by
\begin{equation}
  \label{Eqn:DMD5}
  \chi_n^2 :=  \left \langle \left \Vert \bm{r}_{i}^m
    \right \Vert^2_{\Omega} \right \rangle_M .
\end{equation}
We then select the mode $i$ with the lowest averaged error, i.e.,
\begin{equation}
  \label{Eqn:DMD6}
  \bm{u}_i :=  \bm{u}_n^{\star} \quad \hbox{s.t.} \qquad
  \forall l \in \{1,\ldots,M\} \colon \chi_n^2 \le \chi_l^2,
\end{equation}
and the resulting expansion, after the $i$-th step, reads
\begin{equation}
  \label{Eqn:DMD7}
  \bm{v}_r :=  \sum\limits_{j=1}^i a_j \bm{u}_j + \rm{r}_i^m.
\end{equation}

The $(i+1)$-th mode is computed following the same steps. The
iteration terminates when the desired number of modes is reached,
$i=N$, or in the unlikely case that all residuals vanish $\bm{r}_i^m
\equiv 0, \quad m=1,\ldots, M$ --- whichever criterion is satisfied
first.

%Figure \ref{fig:podmd_algorithm} depicts the algorithm.
%\begin{figure}
%  \centerline{\includegraphics[height=7cm,width=13cm]{PODMD_alg.eps}}
%  \centerline{\includegraphics[height=7cm]{images/PODMD_alg}}
%  \caption{Block diagram of the proposed algorithm}
%\label{fig:podmd_algorithm}
%\end{figure}
%23456789012345678901234567890123456789012345678901234567890123456789012
%-----------------------------------------------------------------------
\section{Modal decomposition of the transient cylinder wake}
\label{ToC:Results}
%-----------------------------------------------------------------------

The transient cylinder wake is analysed with POD
(\S~\ref{ToC:Results:POD}), DMD (\S~\ref{ToC:Results:DMD}) and the
proposed new decomposition (\S~\ref{ToC:Results:RDMD}) discussed in
\S~\ref{ToC:Method}. In \S~\ref{ToC:Results:Comparison} all modal
decompositions are subjected to a comparison with respect to the
instantaneous residual, the averaged residual and the convergence with
increasing number of modes. The transient wake snapshots are the same
for all decompositions and have been described in \S~\ref{ToC:DNS}.

%-----------------------------------------------------------------------
\subsection{Proper Orthogonal Decomposition (POD)}
\label{ToC:Results:POD}

%----- Figure ----------------------------------------------------------
\begin{figure}
  \begin{centering}
    \includegraphics[width=0.75\textwidth]{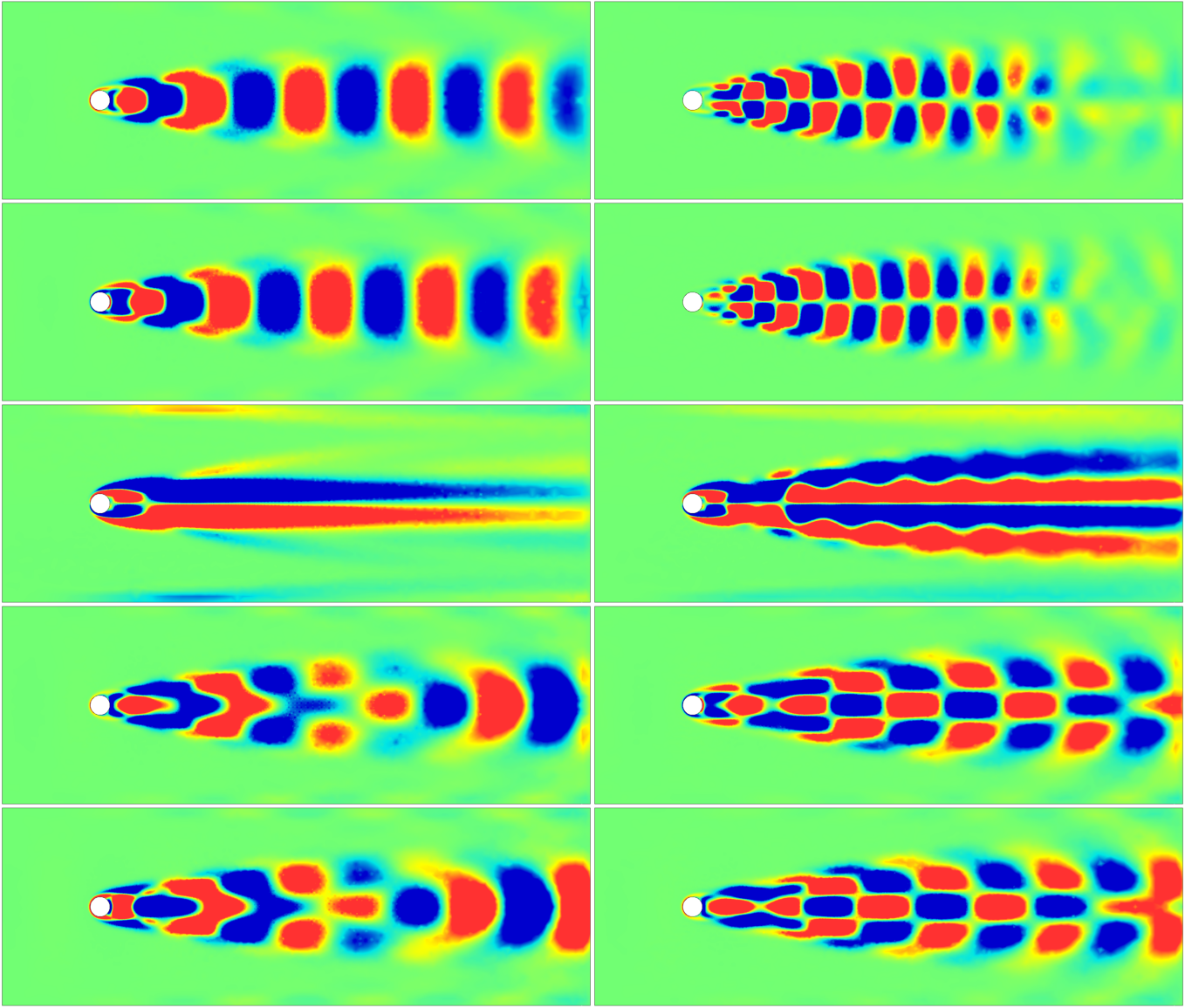}
    \par\end{centering}
  \protect\caption{The first ten POD modes of the flow's transient
    phase with $t \in [30,120]$ and $Re=100$. The first column
    contains the modes $1$--$5$ (top to bottom) and the second column
    modes $6$--$10$ (top to bottom). The vorticity of the mode is
    visualized in color (green: zero, red: above a positive threshold,
    blue: below a negative threshold).}
  \label{fig:pod_modes}
\end{figure}
%----- End of figure ---------------------------------------------------

The snapshots of the cylinder wake simulations described in
\S~\ref{ToC:DNS} are subjected to a snapshot Proper Orthogonal
Decomposition (POD) outlined in \S~\ref{ToC:Method:POD}. The POD modes
of the post-transient \emph{periodic} cylinder wake mimic a Fourier
decomposition. They arise in pairs representing the two phases at the
first and higher harmonic frequencies
\citep{Deane1991pfa,Noack2003jfm}. The energy level of each pair
rapidly decreases with the order of the harmonics. The transient,
however, exhibits a gradual change from the initial stability modes
(with a fluctuation maximum far from the cylinder) to von K\'arm\'an
vortex shedding (with fluctuations peaking near the cylinder). The
other change is a base flow with a recirculation region which
decreases in streamwise extent from about $7$ diameters to about $1$
diameter length. The resulting POD modes and associated mode
amplitudes are depicted in figures \ref{fig:pod_modes} and
\ref{fig:pod_coefficients} and appear different from the
post-transient analogs. POD modes 1 and 2 effectively represent von
K\'arm\'an vortex shedding growing from zero to asymptotic values.
These values are larger than the corresponding fluctuation levels of
the other modes. Modes 6 and 7 describe the second harmonics as can be
seen from the visualisation and the amplitude evolution. Modes 9 and
10 look similar to the stability eigenmodes at slightly lower
frequencies with have a peak activity around $t=55.$ Modes 4 and 5
also represent vortex shedding with a peak activity around $t=65$,
i.e.\ two shedding periods later. Mode 3 depicts the shift mode
\citep{Noack2003jfm}, i.e.\ it characterises the base flow change
between steady and time-averaged periodic solution. Mode 7 represents
another base flow correction with different topology and is mainly active
during the most rapid changes of the transient around $t=65$. It has a
small second harmonic component.

%----- Figure ----------------------------------------------------------
\begin{figure}
  \begin{centering}
    \includegraphics[width=0.75\textwidth]{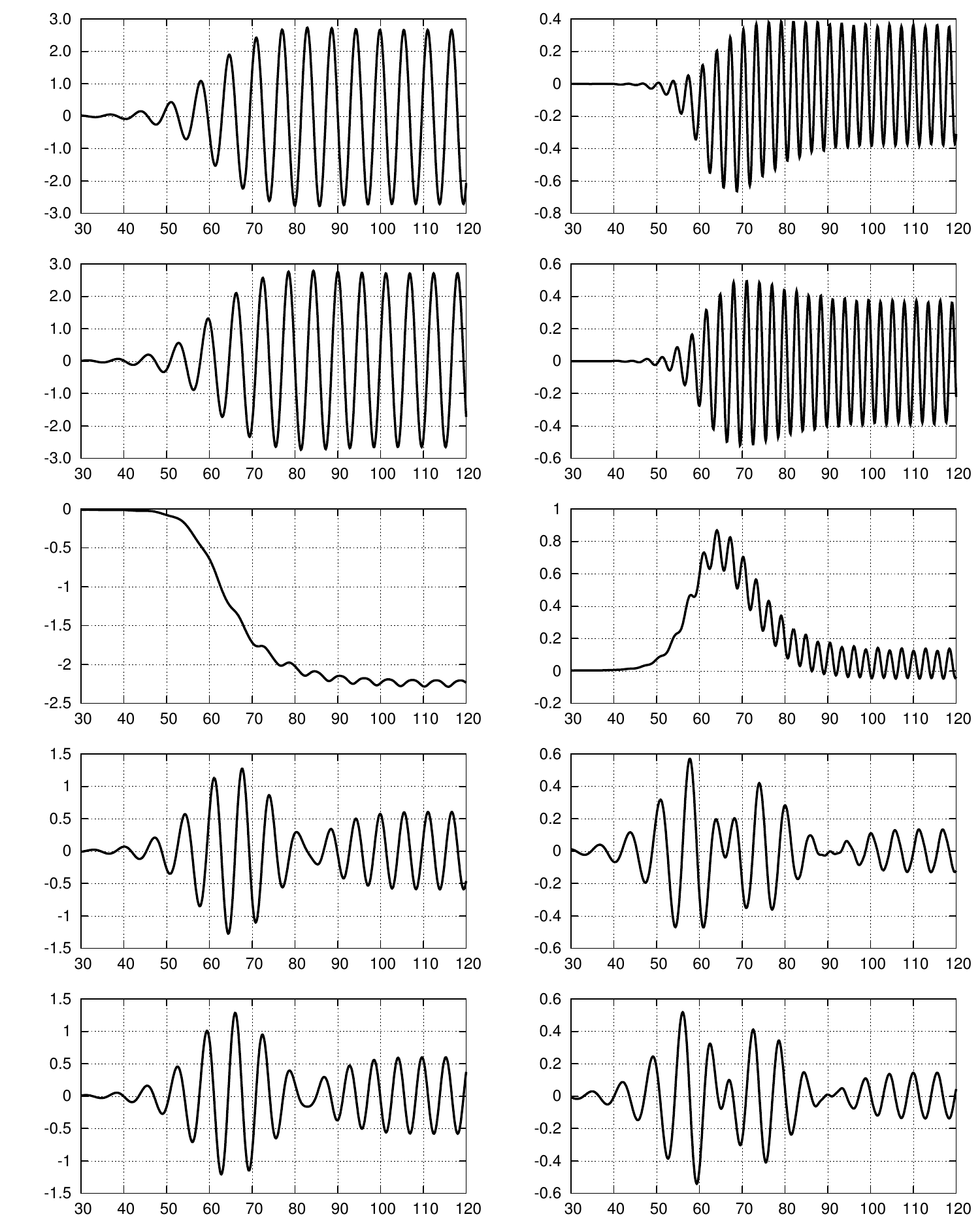}
    \par\end{centering}
  \protect\caption{The mode amplitudes of the POD modes of figure
    \ref{fig:pod_modes} for the corresponding column and row.}
  \label{fig:pod_coefficients}
\end{figure}
% ----- End of ----------------------------------------------------------

All displayed POD modes show pronounced frequencies: six modes display
oscillations near the shedding frequency, two resolve the second
harmonics and two feature slowly varying base-flow changes. Unlike POD
for periodic shedding, there are no traces of the third and higher
harmonics in the first $10$ modes. This comparatively 'clean'
frequency content may be attributed to the narrow-bandwidth transient
dynamics, with dominant frequencies between the eigenfrequency of the
steady solution and the shedding frequency of the post-transient
wake. Moreover, the maximum of the fluctuation envelope moves upstream
during the transient. For broadband dynamics, such as a mixing layer
with multiple vortex pairings, POD modes with multiple frequency
content are far more common \citep{Noack2004swing}.

%-----------------------------------------------------------------------
\subsection{Dynamic Mode Decomposition (DMD)}
\label{ToC:Results:DMD}

The snapshots of the transient flow have been decomposed
with the Dynamic Mode Decomposition.
The DMD procedure can identify eigenmodes for the transient data
and the Fourier modes for the attractor.

%----- Figure ----------------------------------------------------------
\begin{figure}
  \begin{centering}
    \includegraphics[width=0.75\textwidth]{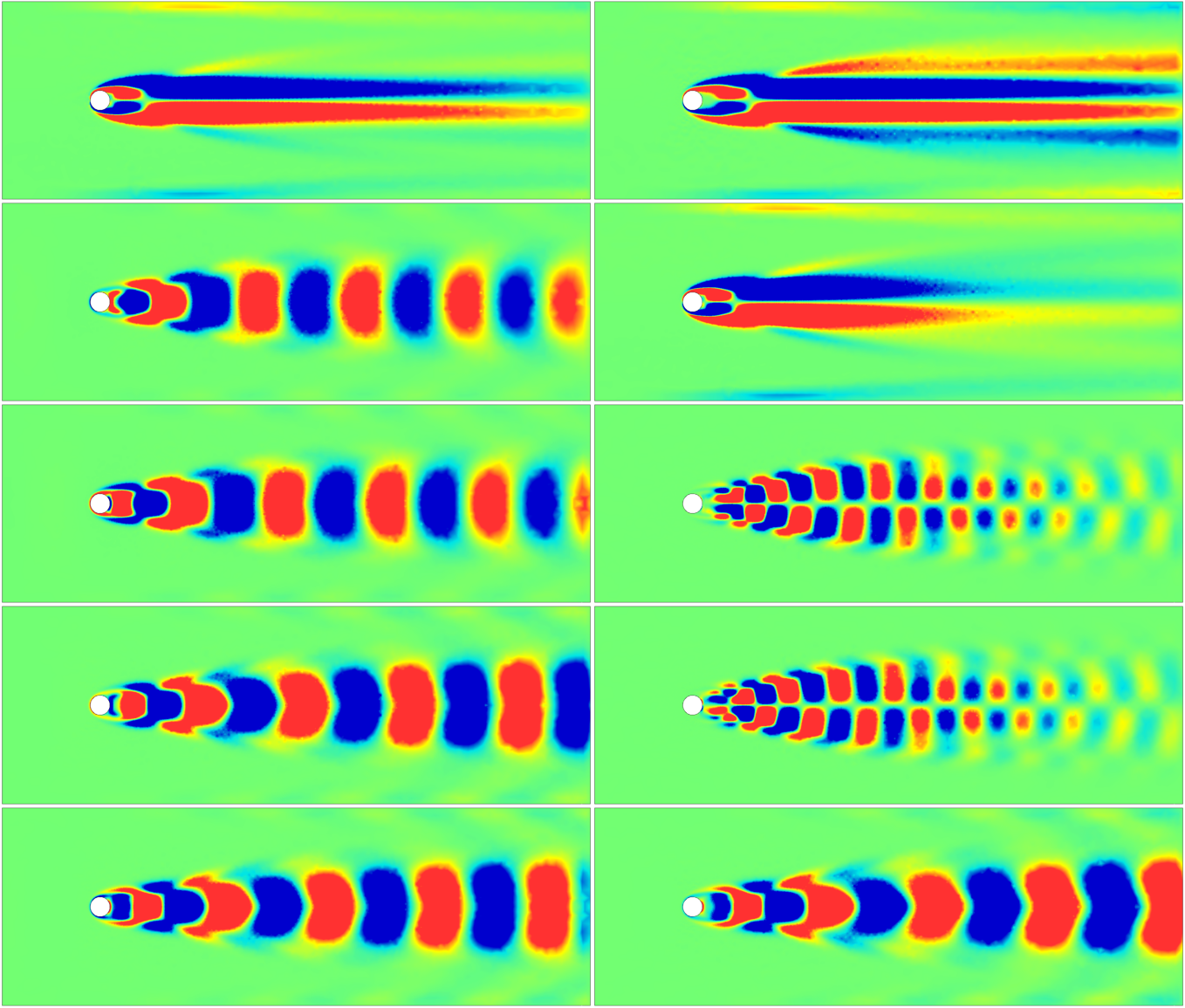}
    \par\end{centering}
  \protect\caption{Same as figure \ref{fig:pod_modes} but for the
    first ten DMD modes.}
  \label{fig:dmd_modes}
\end{figure}
%----- End of figure ---------------------------------------------------

%----- Figure ----------------------------------------------------------
\begin{figure}
  \begin{centering}
    \includegraphics[width=0.75\textwidth]{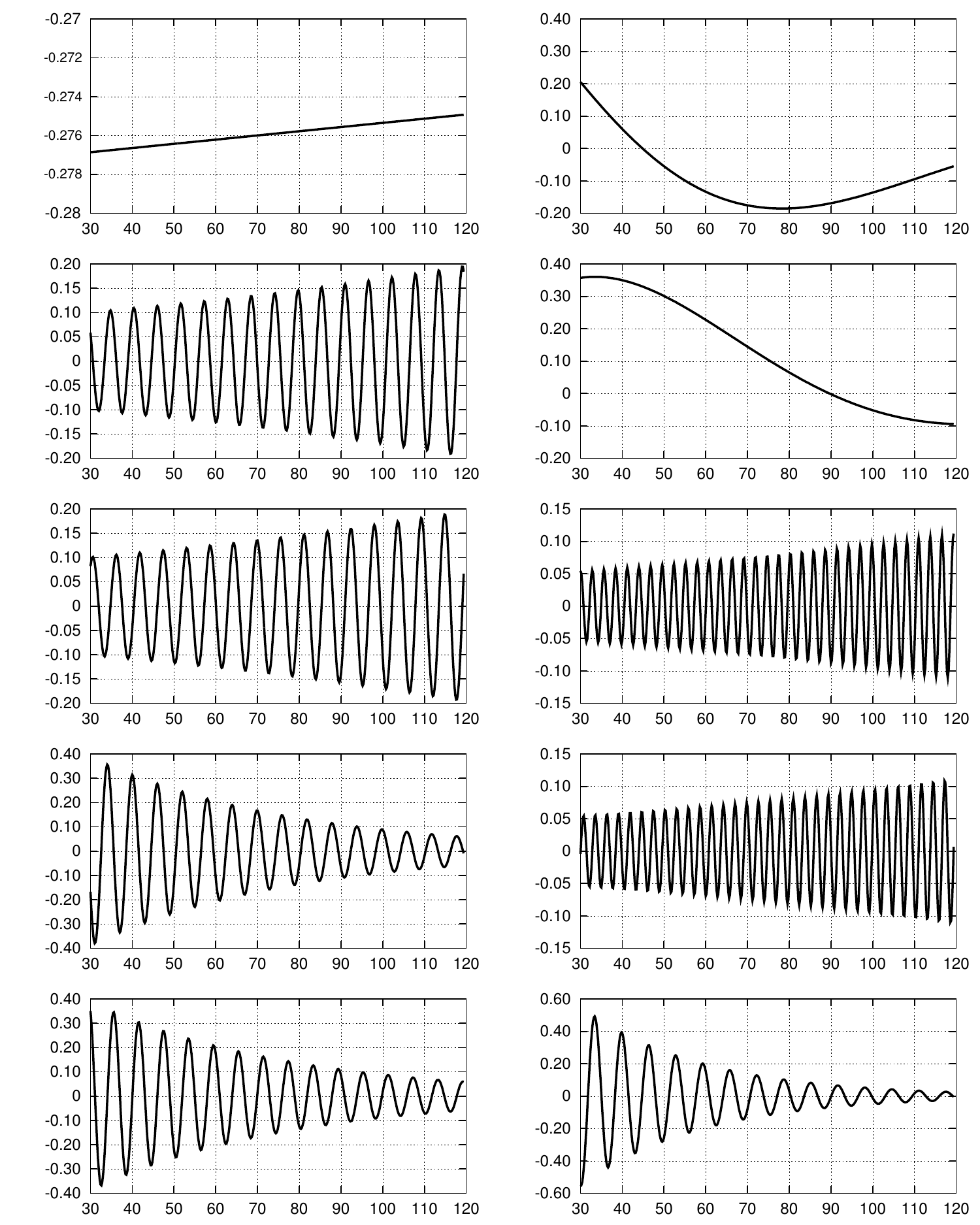}
    \par\end{centering}
  \protect\caption{Same as figure \ref{fig:pod_coefficients} but for
    the first ten DMD modes.}
  \label{fig:dmd_coefficients}
\end{figure}
%----- End of figure ---------------------------------------------------

The result of the procedure is a set of complex Ritz vectors and
complex eigenvalues characterizing the growth rate and the frequency
of the respective mode. The first ten eigenmodes are depicted in
figure \ref{fig:dmd_modes}. The first DMD mode corresponds to a real
eigenvalue leading to a real Ritz vector. The remaining modes
represent the real and imaginary parts of complex Ritz vectors. The
phase-shifted analog of the oscillatory mode 10 is mode 11 (not
displayed). In figure \ref{fig:dmd_coefficients} the corresponding
amplitudes of the real modes are displayed.

Modes 1, 6 and 7 act as shift modes and resolve slow base-flow
changes. This interpretation is corroborated by the behaviour of the
mode amplitudes. The remaining modes describe vortex shedding
($i=2,3,4,5,10$) or its second harmonics ($i=8,9$). The oscillatory
mode amplitudes are slowly growing, like the first vortex shedding
pair (for $i=2,3$) and the second harmonics (for $i=8,9$), or slowly
decaying (for $i=4,5,10$). Intriguingly, the DMD modes describing
vortex shedding have nearly identical frequencies and nearly identical
shapes. This implies a redundancy which constitutes a challenge for
reduced-order modeling.

By construction, the mode amplitudes have an exponentially growing or
decaying envelope and can thus give no indication of initial or
asymptotic values or temporal periods of maximum activity. In
particular, an extrapolation beyond the sampling interval 
$t \in [30,120]$ is not meaningful. An additional potential challenge for
reduced-order modeling is posed by the non-orthogonality of the DMD
modes.

Already in the early literature
\citep{Rowley2009jfm,Schmid2010jfm,Chen2012jns}, DMD has been shown to
accurately capture the onset of fluctuations in the linear regime or
the post-transient behaviour on the attractor. The current results
indicate difficulties of the DMD concerning the modal interpretation
for a complete transient from the steady solution to the
post-transient attractor. This issue has also been pointed out and
analyzed by \citet{Bagheri2013jfm}. In the next section, we present an
alternative DMD decomposition that addresses and removes the
above-mentioned challenges.

%-----------------------------------------------------------------------
\subsection{Recursive Dynamic Mode Decomposition (RDMD)}
\label{ToC:Results:RDMD}

In this section, the results of RDMD are presented. The procedure is
demonstrated for the same set of snapshots as employed previously for
POD and DMD.

%----- Figure ----------------------------------------------------------
\begin{figure}
  \begin{centering}
    \includegraphics[width=0.75\textwidth]{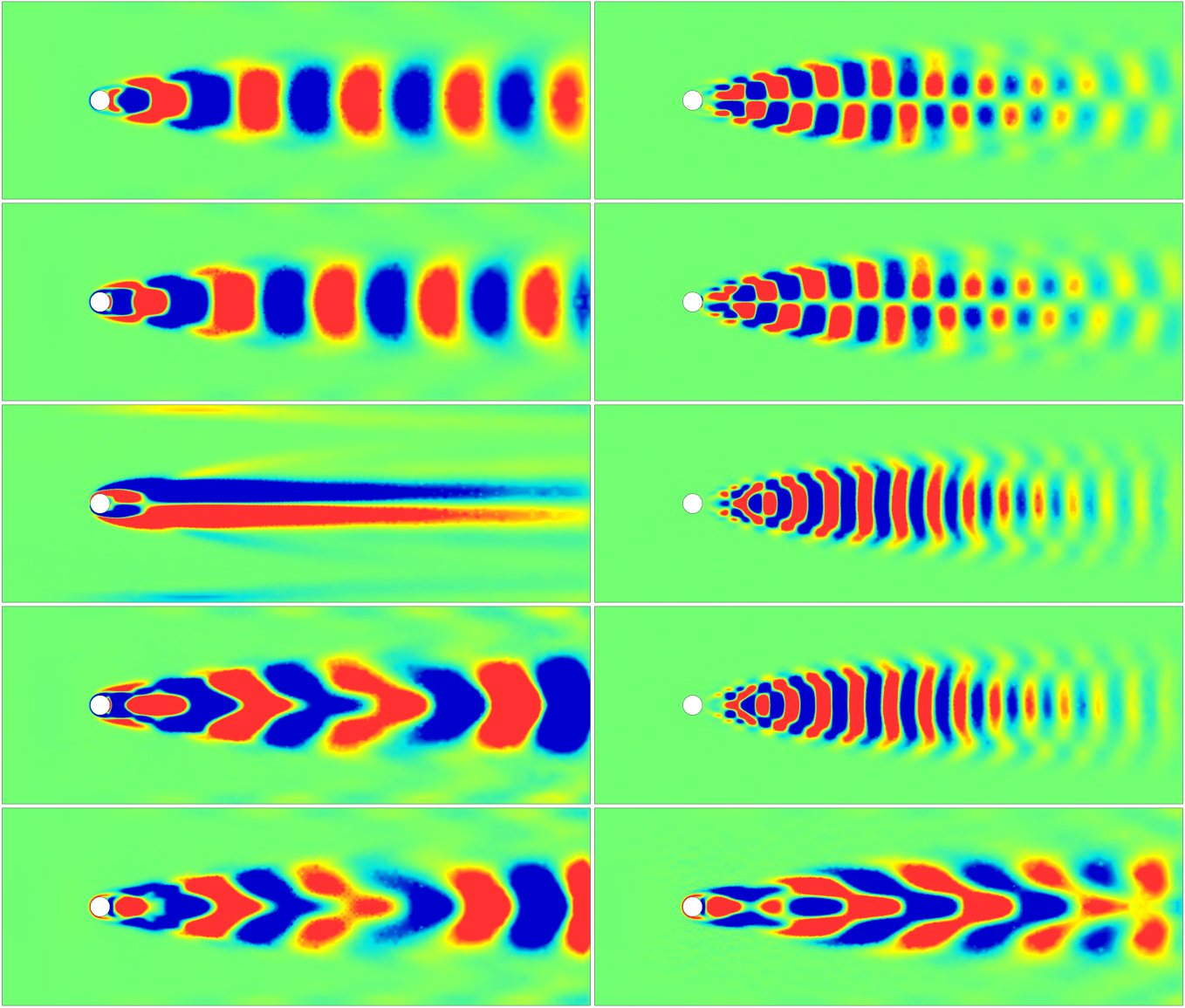}
    \par\end{centering}
  \protect\caption{Same as figure \ref{fig:pod_modes} but for the
    first ten RDMD modes.}
  \label{fig:rdmd_modes}
\end{figure}
%----- End of figure ---------------------------------------------------

%----- Figure ----------------------------------------------------------
\begin{figure}
  \begin{centering}
    \includegraphics[width=0.75\textwidth]{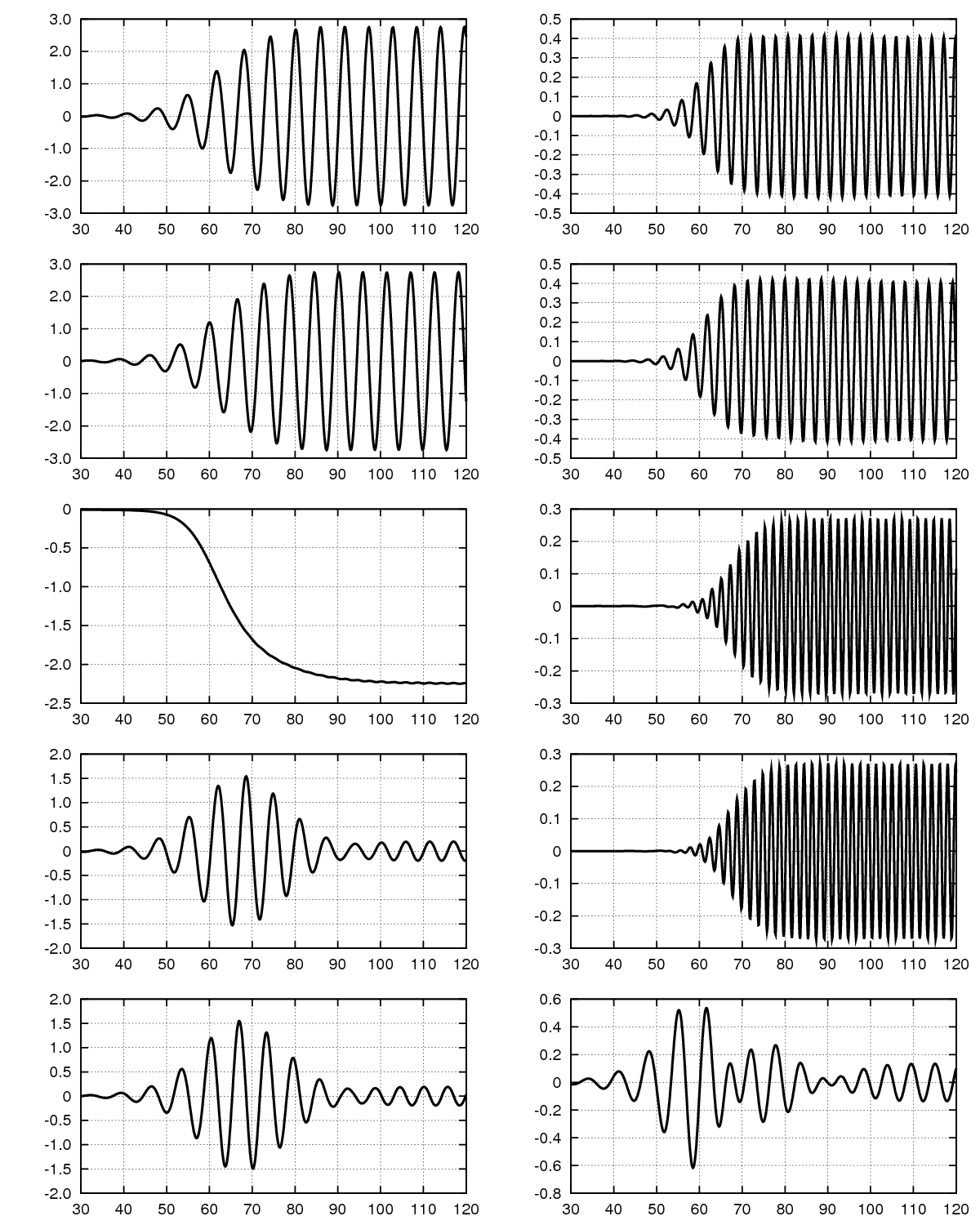}
    \par\end{centering}
  \protect\caption{Same as figure \ref{fig:pod_coefficients} but for
    the first ten RDMD modes.}
  \label{fig:rdmd_coefficients}
\end{figure}
%----- End of figure ---------------------------------------------------

The first ten RDMD modes are depicted in figure \ref{fig:rdmd_modes}.
Intriguingly, RDMD resolves the first harmonics (modes $i=1,2$), the
second harmonics ($i=6,7$) and the third harmonics ($i=8,9$). The
associated mode amplitudes show corresponding oscillations starting
near zero and reaching asymptotic values on the limit cycle. Mode 3
shows a nearly pure shift mode, describing the transition of the base
flow from the steady solution to the time-averaged flow. In contrast
to \citet{Noack2003jfm}, the amplitude becomes negative, since the
sign of the RDMD mode is arbitrary. Modes 4 and 5 resolve intermediate
vortex shedding patterns with maximum activity around $t=70$ and
rather small residual fluctuations on the limit cycle. Modes 10 and 11
(not shown) are reminiscent of stability eigenmodes and peak near
$t=55$, i.e.\ more than two periods earlier.

The first seven modes have significant similarities with the POD modes
of figures \ref{fig:pod_modes} and \ref{fig:pod_coefficients}.  Yet,
the oscillations are more symmetric (compare RDMD and POD mode 6) and
show no apparent frequency mixing, as in POD modes $i=3,6,8$. RDMD
modes have by definition a lower averaged residual, as compared to POD
modes, but they look much cleaner and even reveal the third harmonics.
It appears that the RDMD modes have more in common with POD than with
DMD modes. This should not come as a surprise, since the primary
construction principle is the minimization of the residual while the
secondary criterion is the emulation of single-frequency DMD-mode
behaviour.

%-----------------------------------------------------------------------
\subsection{Comparison of POD, DMD and RDMD}
\label{ToC:Results:Comparison}

In this section, results from POD, DMD and RDMD are quantitatively
compared.

%----- Figure ----------------------------------------------------------
\begin{figure}
  \begin{centering}
    \includegraphics[width=0.75\textwidth]{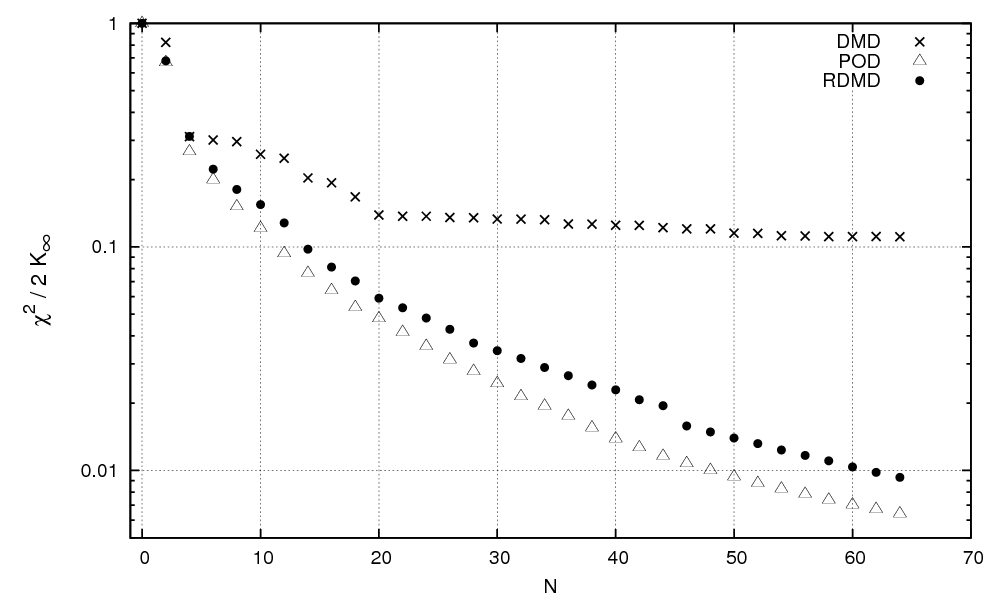}
    \par\end{centering}
  \protect\caption{The time-averaged fluctuation level of the residual 
  \eqref{Eqn:TruncationError}
    for POD, DMD and RDMD
    with increasing number of modes.  
    The value is normalized by
    the corresponding fluctuation level ($2K_{\infty}$) on the limit cycle.}
  \label{fig:decrease_of_residuum_with_modes}
\end{figure}
% ----- End of figure ---------------------------------------------------

%----- Figure ----------------------------------------------------------
\begin{figure}
  \begin{centering}
    \includegraphics[width=0.75\textwidth]{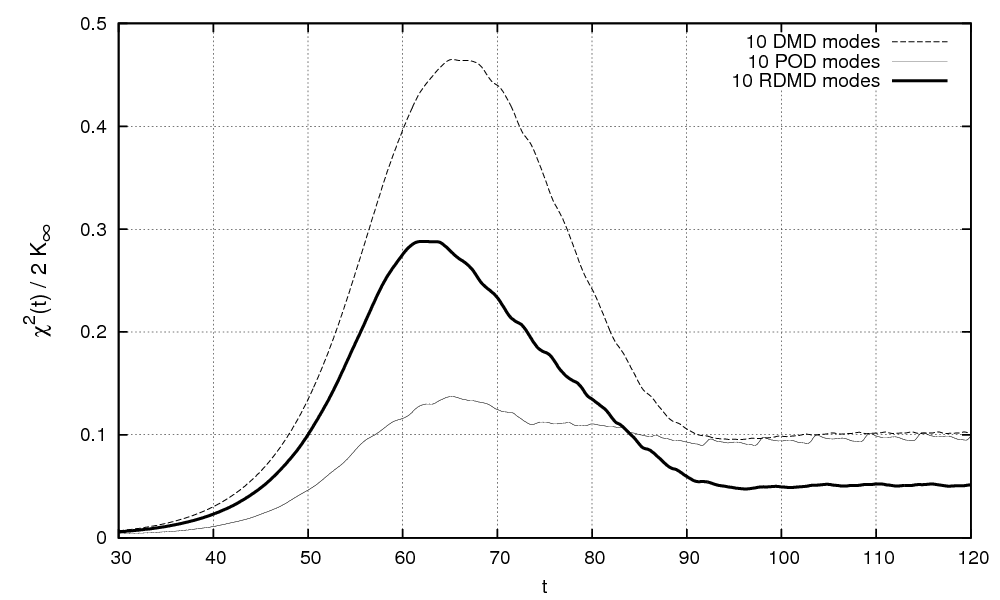}
    \par\end{centering}
  \protect\caption{The instantaneous normalized truncation error \eqref{Eqn:TruncationError} 
    as a function of time
    for 10 POD, DMD and RDMD modes. The figure displays the sampling
    interval.}
  \label{fig:10-mode_transient}
\end{figure}
%----- End of figure ---------------------------------------------------

%----- Figure ----------------------------------------------------------
%\begin{figure}
%\begin{centering}
%\includegraphics[width=0.75\textwidth]{TKE-ROM-cmp1.png}
%\par\end{centering}
%\protect\caption{Galerkin model build with 10 POD, DMD and RDMD modes.
%\textcolor{red}{TRY HIGHER INITIAL FLUCTUATION LEVEL
%SO THAT THE START OF THE TRANSIENT IS COMPARABLE.}}
%\label{fig:Galerkin_model}
%\end{figure}
%----- End of figure ---------------------------------------------------

Figure \ref{fig:decrease_of_residuum_with_modes} shows the truncation error of
the POD, DMD and RDMD expansions as a function of the number of modes.
As expected, all errors decrease monotonically with the number of
modes and POD outperforms DMD and RDMD. The RDMD residual, however,
follows the POD value remarkably well and stays within similar orders
of magnitude. In contrast, DMD approaches an effective asymptote near
$10\%$ of the final fluctuation level.

The temporal evolution of the instantaneous truncation error is displayed in
figure \ref{fig:10-mode_transient} for POD, DMD and RDMD at $N=10$.
The maximum value in $t \in [60,70]$ is lowest for POD and largest
for DMD. RDMD performs, as expected, between the alternative
expansions. A surprising feature are the asymptotes. POD and DMD have
similar truncation errors near $10\%$ of the terminal fluctuation level while
RDMD outperforms both with a final value at about $5\%.$ This is no
contradiction to the optimal property of POD, as this property only
guarantees a minimal \emph{time-averaged} value, or, equivalently, a
minimal value of the integral over the instantaneous truncation error, which it evidently
has. The DMD-based frequency filtering included in RDMD appears to
have 'anticipated' the limit cycle behaviour. One indication of this
'anticipation' is the third harmonics which is featured in RDMD
but absent in both POD and DMD. At this stage, RDMD appears more
suitable for reduced-order modeling when compared to POD or DMD.  Like
POD, RDMD yields orthonormal modes, but with purer frequency content.
Like DMD, RDMD extracts oscillatory modes, but with well-defined
initial and asymptotic behaviour of the mode amplitudes.

%--------------------------------------------------------------------------
\section{Conclusions}
\label{ToC:Conclusions}
%--------------------------------------------------------------------------

We propose a novel data-driven flow decomposition which combines the
modal orthonormality and low truncation error $\chi^2$ 
\eqref{Eqn:TruncationError}
of POD with the frequency-distilling features of DMD. This decomposition is
recursively defined. First, the data set is subjected to DMD, after
which a normalized DMD mode is chosen which minimizes the averaged
error $\chi^2$. This procedure is recursively repeated in the orthogonal
subspace of computed modes. The resulting \emph{recursive DMD (RDMD)
  modes} are orthonormal by construction and can be expected to have
lower error $\chi^2$ than DMD modes while retaining the monochromatic
features of DMD.

POD, DMD and RDMD have been applied to the same snapshots of a
transient cylinder wake starting near the steady solution and
terminating on the limit cycle. As expected, RDMD significantly
outperforms DMD in terms of the maximum, time-averaged and asymptotic
truncation error for all considered mode numbers by a large margin. In
addition, the exponentially growing or decaying DMD amplitudes neither
resemble initial nor asymptotic flow behaviour in a meaningful manner
while RDMD amplitudes clearly identify initial, transient and
post-transient flow phases.

%----- Table -----------------------------------------------------------
\begin {table}
  \label{table:summary}
  \caption{Comparison of POD, DMD and RDMD for the transient cylinder wake.
The maximum and asymptotic truncation errors are 
for expansions with $N=10$  modes normalized with the post-transient fluctuation level $2 K_{\infty}$.}
  \begin{center}
    \begin{tabular}{l|c|c|c}
      % .......................................................................
      & \rule{25mm}{0pt}
      & \rule{25mm}{0pt}
      & \rule{25mm}{0pt}
      \\[-10pt]
      Aspect
      & POD
      & DMD
      & RDMD\rule[-10pt]{0pt}{25pt}
      % .......................................................................
      \\ \cline{1-4}
      Ideal snapshots
      & uncorrelated
      & time-resolved
      & time-resolved\rule[-10pt]{0pt}{25pt}
      % .......................................................................
      \\ \cline{1-4}
      Averaged truncation error
      &  optimal
      &  poor
      &  good\rule[-10pt]{0pt}{25pt}
      % .......................................................................
      \\ \cline{1-4}
      Maximum truncation error 
      &  13\% 
      &  46\%
      &  28\% \rule[-10pt]{0pt}{25pt}
      % .......................................................................
      \\ \cline{1-4}
      Asymptotic truncation error
      &  10\% 
      &  10\% 
      &   5\% \rule[-10pt]{0pt}{25pt}
      % .......................................................................
      \\ \cline{1-4}
      Modal frequency content
      & mixed
      & pure
      & almost pure\rule[-10pt]{0pt}{25pt}
      % .......................................................................
      \\ \cline{1-4}
      Noise sensitivity
      & low
      & high
      & like\rule[-5pt]{0pt}{20pt}
      \\
      &
      & without filter
      & DMD\rule[-10pt]{0pt}{20pt}
      % .......................................................................
      \\ \cline{1-4}
      Niche applications
      & statistics
      & stability modes
      & transient\rule[-5pt]{0pt}{20pt}
      \\
      &
      & Fourier modes
      & dynamics\rule[-10pt]{0pt}{20pt}
    \end{tabular}
  \end{center}
\end{table}
% -----------------------------------------------------------------------

Also as expected, RDMD modes have far purer frequency content than POD
but maintain the residual at comparable level. While the maximum and
average truncation error of POD outperforms RDMD, RDMD shows a better
resolution of the limit cycle: the asymptotic value for $N=10$ is
half as large as the corresponding POD value and the third harmonic
frequency is only captured by RDMD. Table \ref{table:summary} provides
a brief comparison of the main characteristics and features of POD,
DMD and RDMD.

The literature contains alternative approaches for spectrally purified
POD modes. One important recent contribution is spectral POD
\citep{Sieber2015arXiv} which interpolates between POD and DMD by
filtering the correlation matrix. This continuous interpolation offers
an additional degree of freedom not present in RDMD; the price is the
loss of strict orthonormality of the modes for all interpolation
parameters. In a similar vein, \citet{Bourgeois2013jfm} construct
orthogonal POD modes with purer frequency content after
Morlet-filtering the flow data at design frequencies. RDMD can be
considered as a simpler approach which can be expected to perform well
in an unsupervised manner, but leaves little room for tuning the modes
for special purposes or applications.

The current study indicates that RDMD is an attractive 'compromise'
between POD and DMD. It sacrifices little of the optimal residual
property of POD while retaining the single-frequency behaviour of DMD.
Its potential in control-oriented reduced-order modeling will be
explored in a future effort. In addition, a particularly attractive
opportunity for RDMD is the unsupervised extraction of generalized
mean-field models with few dominant frequencies
\citep{Brunton2015amr}.\section*{Acknowledgements}
B.N. acknowledges the funding and excellent working conditions 
of the Senior Chair of Excellence
'Closed-loop control of turbulent shear flows 
using reduced-order models' (TUCOROM 2010--2015)
supported by the French Agence Nationale de la Recherche (ANR)
and hosted by Institute PPRIME
and the Collaborative Research Centre (CRC 880) 
`Fundamentals of High Lift of Future Civil Aircraft'
supported by the Deutsche Forschungsgemeinschaft (DFG) 
and hosted at the Technical University of Braunschweig.
This work has also been funded by National Centre of Science under research
grant no. DEC-2011/01/B/ST8/07264 ''Novel method of Physical Modal Basis
Generation for Reduced Order Flow Models''.
We have highly profited from stimulating discussions 
with Markus Abel, Steven Brunton, Laurent Cordier, 
Robert Martinuzzi, Bartosz Protas, Rolf Radespiel and Richard Semaan.
We thank the Ambrosys Ltd.\ Society for Complex Systems Management,
the Bernd Noack Cybernetics Foundation for additional support.
%\end{acknowledgement}

%\begin{appendix}
%\input{sA.tex}
%\end{appendix}

\end{document}